\documentclass[12pt]{article}
\begin{document}
\begin{center}
{\bf Particle production by the thick-walled bubble}

\vspace{0.8cm} { Michael Maziashvili }

\vspace{0.5cm}\baselineskip=14pt

\vspace{0.5cm} \baselineskip=14pt {\it Department of Theoretical
Physics, Tbilisi State University, 0128 Tbilisi, Georgia}
\end{center}
\vspace{0.5cm}

\begin{abstract}
The spectrum of created particles during the tunneling process,
leading to the decay of a false vacuum state, is studied
numerically in the thick-wall approximation. It is shown that in
this case the particle production is very intensive for small
momenta. The number of created particles is nearly constant
$n(p)\approx 1$ for $4\leq p\leq500$.
\end{abstract}
\section{Introduction}
In the semiclassical approximation, false vacuum decay through the
barrier penetration is described by the nontrivial $O(4)$
symmetric solution of Euclidean (imaginary-time) equation of
motion with the boundary condition that the field asymptotically
approaches its false vacuum value, \cite{Col}. This solution is
referred to as the bounce one. The bounce looks like a
four-dimensional spherical bubble of true vacuum embedded in a
false vacuum background. The bubble evolution after nucleation is
obtained by the analytic continuation of the bounce solution to
the Minkowskian time, $\tau\rightarrow it$. There exist two
limiting cases, referred to as the thin and thick wall
approximations respectively, when the theory of bubble formation
is greatly simplified, \cite{Col,Lin}. Thin wall approximation is
realized when the energy-density difference between the true and
false vacuum is much smaller than the height of the barrier. The
particle production in this case has been studied in \cite{Ma1}.
The analysis presented in that paper shows that the particle
production is strongly suppressed in the thin-wall approximation.
It is of interest to explicate whether this is the general feature
for the created particles spectrum during the false vacuum
tunneling when the background space-time curvature is neglected
(without gravity). Throughout this paper we shall consider another
limiting case, when the energy-density difference between the true
and false vacuum is much grater than the height of the barrier.
The paper is organized as follows. In Sec.2 we briefly describe
the technique for calculation of the particle spectrum. In Sec.3,
based on the calculational procedure described in Sec.2, we
evaluate the particle spectrum for the thick-walled bubble. Sec.4
is devoted to the summary and concluding remarks.

\section{Technique of calculation}

For the calculation of particle spectrum we use the standard
technique developed in \cite{YTS,HSTY}. Throughout this paper the
metric signature in Minkowski space-time is $(-,+,+,+)$ and units
$\hbar=c=1$ are used. We consider a scalar field $\phi$ defined by
the Lagrange density \begin{equation}{\cal L}
=-\frac{1}{2}(\partial_{\mu}\phi)^2-U(\phi),\end{equation} with
the potential
\begin{equation}U(\phi)=\frac{\lambda}{2}\phi^2(\phi-2a)^2-
\epsilon(\phi^3/2a^3-3\phi^4/16a^4),\label{po}\end{equation} where
$\lambda,~a$ and $\epsilon$ are positive parameters. The potential
(\ref{po}) has a local minimum at $\phi_f=0,~U(0)=0,$ a global
minimum at $\phi_t=2a,~U(2a)=-\epsilon,$ and a local maximum at
$\phi_{top}=8\lambda a^5/(8\lambda
a^4+3\epsilon),~U(\phi_{top})=128\lambda^3 a^{12} (2\lambda
a^4+\epsilon)/(8\lambda a^4+3\epsilon)^3$. The $O(4)$ symmetric
bounce solution $\phi_{b}$ satisfies the Euclidean equation of
motion
\begin{equation}\frac{d^2\phi}{d\rho^2}+\frac{3}{\rho}\frac{d\phi}{d\rho}=U'(\phi),\end{equation}
with the boundary conditions $\phi(\infty)=0,~d\phi/d\rho=0$ at
$\rho=0$, where $\rho=\sqrt{\vec{x}\,^2+\tau^2}$ and the prime
denotes differentiation with respect to $\phi$. Since the bounce
solution is $O(3,1)$ invariant in the Minkowskian space-time and
$O(4)$ invariant in the Euclidean region it is convenient to use
the following coordinate systems. The coordinate system used in
the Euclidean region is $(\rho,\chi,\theta,\varphi)$ where
$(\theta,\varphi)$ are usual angle coordinates on two-dimensional
sphere and $(\rho,\chi)$ are related to $r=|\vec{x}|$ and $\tau$
as follows
\begin{eqnarray}r=\rho\sin(\chi),~\tau=-\rho\cos(\chi),\nonumber \\0\leq\chi\leq
\pi/2,~0\leq\rho<\infty.\end{eqnarray} The coordinate system in
the Minkowskian region may by obtained by replacement $(\rho,\chi)
\rightarrow \\(-i\rho_{M},-i\chi_{M})$, which yields
\begin{eqnarray}r=\rho_{M}\sinh(\chi_{M}),~t=\rho_{M}\cosh(\chi_{M}),\nonumber
\\0<\chi_{M}<\infty,~0<\rho_{M}<\infty.\end{eqnarray} The equation governing the
fluctuation field $\Phi$ reads
\begin{equation}\left[-\partial_{M}^2-\frac{3}{\rho_{M}}\partial_{M}+\frac{1}{\rho_{M}^2}\hat{L}^2-U''(\phi_b)\right]\Phi=0,\label{1}\end{equation}
where $\partial_{M}$ denotes the partial derivative with respect
to $\rho_{M}$ and $\hat{L}$ is the Laplacian operator on
three-dimensional unit hyperboloid. Expanding $\Phi$ in terms of
harmonic functions on the three-dimensional hyperboloid $Y_{plm}$,
$-\hat{L}^2Y_{plm}=(1+p^2)Y_{plm}$, the Eq.(\ref{1}) for the mode
function takes the form
\begin{equation}\label{al}\left[\partial_{M}^2+\frac{(p^2+1/4)}{\rho_{M}^2}+U''(\phi_{b})\right]\psi=0.\end{equation}
Making use of analytic continuation of $\psi$ to the Euclidean
region by replacement $\rho_{M}\rightarrow i\rho$ one obtains the
equation for the mode function in the under-barrier region
\begin{equation}\label{fo}\left[\partial_{\rho}^2+\frac{(p^2+1/4)}{\rho^2}-U''(\phi_{b})\right]\psi_E=0,\end{equation}
where $\psi_E$ is connected with $\psi$ by the asymptotic boundary
condition $\psi_E(-i\rho_M)=\psi(\rho_M)$ at $\rho_M\rightarrow
0$. At $\tau=-\infty$ the field $\phi$ is in false vacuum state
and correspondingly the fluctuation field satisfies the vanishing
boundary condition $\psi_E\rightarrow 0$ when
$\rho\rightarrow\infty$. Since $U''(\phi_b)\rightarrow 4a^2\lambda
$ when $\rho\rightarrow\infty$ the asymptotic solution of
Eq.(\ref{al}) will have the form
\begin{equation}\psi(\rho_{M})=c_1(p)\sqrt{\rho_{M}}\,H^{(1)}_{ip}(2a\sqrt{\lambda}\rho_{M})+
c_2(p)\sqrt{\rho_{M}}\,H^{(2)}_{ip}(2a\sqrt{\lambda}\rho_{M}),\end{equation}
where $H^{(1)}_{ip}(x)$ and $H^{(2)}_{ip}(x)$ are the Hankel
functions of the first and second kinds, respectively. The
spectrum of created particles $n(p)$ is given by \cite{YTS,HSTY}
\begin{equation}\label{sp}n(p)=\frac{e^{-2p\pi}}{||c_1(p)/c_2(p)|^2-e^{-2p\pi}|}.\end{equation} For detailed description of the general
formalism see \cite{YTS,HSTY}.

\section{Particle spectrum}

For the thick-wall case we specify the potential parameters as
follows $a=1~,\lambda=0.5~,\epsilon=100$. For that values of
parameters the ratio $U(\phi_{top})/\epsilon=0.5752022890e-6$. The
corresponding bounce solution obtained by the program code odex
with local tolerance $1.0D-17$ is shown in Fig.1. The detailed
description of this program code is given in \cite{HNW}. Taking
into account that $\phi_b(10)=0.3453646895630613e-08$ and using
the program package Maple 7 one finds that the solution of
Eq.(\ref{fo}) for $\rho\geq10$ satisfying the vanishing boundary
condition when $\rho\rightarrow\infty$ is given by
\begin{equation}\label{Be}\psi_E=a_3\sqrt{\rho}K_{ip}(\sqrt{2}\rho),\end{equation}
where $K_{ip}(x)$ is the modified Bessel function of the second
kind. In order to make the analytic continuation at $\rho=0$ we
construct the analytic solution to Eq.(\ref{fo}) in the vicinity
of $\rho=0$. For this purpose, the power series expansion of
$\phi_b$ about $\rho=0$ is used for evaluating of $U''(\phi_b)$
\begin{equation}\label{ex}U''(\phi_b)=-29.34386061+50.39180321\rho^2-52.71178464\rho^4-21.76409860\rho^6+O(\rho^8).\end{equation}
Since we want to use the expansion (\ref{ex}) for evaluating the
solution of Eq.(\ref{fo}) in the region $0<\rho<0.001$ we omit the
terms higher than quadratic. Inserting this expression into
Eq.(\ref{fo}) and using the program package Maple 7 one obtains
the following solution
\begin{equation}\label{Wh}\psi_E=\frac{a_1}{\sqrt{\rho}}M(1.033421066,0.5\,i\,p,7.09871842\rho^2)+
\frac{a_2}{\sqrt{\rho}}W(1.033421066,0.5\,i\,p,7.09871842\rho^2),\end{equation}
where $M,~W$ denote the Whittaker functions. For $\rho_M\geq10$
the solution of Eq.(\ref{al}) is given by
\begin{equation}\label{asy}\psi(\rho_{M})=c_1(p)\sqrt{\rho_{M}}\,H^{(1)}_{ip}(\sqrt{2}\,\rho_{M})+
c_2(p)\sqrt{\rho_{M}}\,H^{(2)}_{ip}(\sqrt{2}\,\rho_{M}).\end{equation}
Since the particle number depends on the ratio $|c_1(p)/c_2(p)|$
the multiplier $a_3$ in Eq.(\ref{Be}) may be fixed arbitrarily.
The Eq.(\ref{Be}) and its derivative give the initial conditions
to Eq.(\ref{fo}) at $\rho=10$. Using these initial values the
solution of Eq.(\ref{fo}) in the region $0.001\leq\rho<10$ is
constructed numerically by the code odex with local tolerance
$1.0D-17$. Matching this solution with Eq.(\ref{Wh}) at
$\rho=0.001$ one finds the coefficients $a_1,~a_2$. Making use of
analytic continuation $\rho\rightarrow -i\rho$ the Eq.(\ref{Wh})
now gives the initial conditions to Eq.(\ref{al}) at
$\rho_M=0.001$. With these initial conditions the Eq.(\ref{al}) is
solved again numerically up to $\rho_M=10$ by the odex with local
tolerance $1.0D-17$. Matching this solution with Eq.(\ref{asy})
and using Eq.(\ref{sp}) one evaluates the particle number for
given $p$. For evaluating of special functions and their
derivatives at the matching points we have used the program
package Maple 7. The spectrum obtained in this way is shown in
Fig.2. As $p$ increases from $0$ to $4$ the number of created
particles decreases from $23714.66515$ to $1.000916176$ and
becomes nearly constant $n(p)\approx 1$ for $4\leq p\leq 500$. For
the sake of clarity the behavior of $n(p)$ for small values of
momentum is shown in the following table.
\[\begin{tabular}{|l|l|l|l|l|l|}\hline
p & 0.25 & 0.5 & 0.75 & 1 & 1.25  \\\hline n & 15524.29139 &
6622.761821  & 2612.820881 & 1008.019999 & 378.9347624
\\\hline\hline p & 1.5 & 1.75 & 2 & 2.25 & 2.5 \\\hline
n & 138.1763281 & 49.01973850 & 17.20855685 & 6.281386004 &
2.664255183 \\\hline
\end{tabular}\]

\section{Summary}
Using the standard technique \cite{YTS,HSTY} we have shown that,
in contrast to the thin-wall case, the particle production in the
thick-wall approximation is very intensive. The number of created
particles is especially large at small values of momentum,
decreases sufficiently fast as $p$ increases from $0$ to $4$ and
becomes nearly constant $n(p)\approx 1$ for $4\leq p\leq 500$. To
date we have been unable to evaluate the behavior of $n(p)$ for
asymptotically large values of $p$. In the thin-wall case the
particle production is strongly suppressed in general, it is
nearly constant for small momenta and behaves as $\exp(-2p\,\pi)$
for large values of momentum \cite{Ma1}. As we observe the
particle production has a quite different nature in the thick-wall
case. A few remarks are in order. This result is not strictly
correct since we do not take into account the constraint on the
fluctuation field that arises due to fact that the proper
fluctuation field associated with the tunneling is the transverse
part of total fluctuation field with respect to the bounce
solution \cite{Ma2}. In \cite{Ma2} the application of this
constraint is illustrated for the spatially homogenous tunneling.
It was shown that, due to this constraint, the spatially
homogenous tunneling does not allow the particle production with
zero momentum. If such a restriction will not reduce the particle
number significantly in the thick-wall case, one has to consider
the back reaction of particle production on the tunneling. On the
other hand the analysis of a concrete thick-wall model, when the
quartic term is ignored in Eq.(\ref{po}), shows that the
activation rate at zero temperature makes a dominant contribution
to the vacuum decay in comparison with the tunneling rate
\cite{CRV}. Unfortunately we do not know whether this is valid in
the thick-wall approximation in general and if this is the case
how does the zero temperature activation rate depend on
$a^4\lambda/\epsilon$.

\section*{Acknowledgments}

We are grateful for useful conversations with Professors
A.\,Khelashvili and I.\,Lomidze.

\end{document}